\documentclass[lettersize,journal]{IEEEtran}
\usepackage{algorithmic}
\usepackage{array}
\usepackage[caption=false,font=normalsize,labelfont=sf,textfont=sf]{subfig}
\usepackage{textcomp}
\usepackage{stfloats}
\usepackage{url}
\usepackage{verbatim}
\usepackage{graphicx}
\usepackage{amsmath, amssymb, amsfonts, bm, booktabs}
\def\BibTeX{{\rm B\kern-.05em{\sc i\kern-.025em b}\kern-.08em
    T\kern-.1667em\lower.7ex\hbox{E}\kern-.125emX}}
\usepackage{balance}
\usepackage{hyperref, cite}
\usepackage{multirow}
\usepackage{circuitikz}

\usepackage{mathtools}
\usepackage{tcolorbox}  
\usepackage{bm}         
\usepackage{multirow}   
\usepackage{array}
\usepackage{mathrsfs}  
\usepackage{needspace}
\usepackage{tabularx} 
\usepackage{pifont}
\usepackage[intoc]{nomencl}
\usepackage[linesnumbered,ruled,vlined]{algorithm2e}
\usepackage{xcolor}
\usepackage{enumitem}
\usepackage{tabularx}
\usepackage{makecell}
\SetAlgoNlRelativeSize{0}  

\newcommand{\autonumlabel}[1]{%
  \refstepcounter{equation}%
  \label{#1}%
  (\theequation)%
}

\hypersetup{
    colorlinks=true,     
    linkcolor=blue,      
    citecolor=blue,       
    filecolor=magenta,     
    urlcolor=cyan        
}

\ifCLASSINFOpdf
\else
\fi
\begin{document}
%
\title{DCCB-Capability-Aware Optimal Info-Power Flow for Transmission Switching in AC/MTDC Grids}

%
%
%

\author{Haixiao Li,
        Aleksandra Lekić

\thanks{Haixiao Li and Aleksandra Lekić are with the Faculty of Electrical Engineering, Mathematics and Computer Science, TU Delft, 2628 CD Delft, The Netherlands (e-mail: \href{mailto:haixiaoli.ee@gmail.com}{haixiaoli.ee@gmail.com};
\href{mailto:a.lekic@tudelft.nl}{a.lekic@tudelft.nl}).}%
}

%
%

\markboth{ xxx,~Vol.~XX, No.~XX, August~2026}%
{YANG \MakeLowercase{\textit{et al.}}: Bare Demo of IEEEtran.cls for IEEE Journals}
%



\maketitle
\begin{abstract}
Transmission switching is an effective approach for restoring operational flexibility and improving the economic performance of AC/multi-terminal DC (MTDC) grids following DC-side faults. Its practical implementation requires that the faulted branch be safely isolated and that the associated switching commands be successfully delivered through the communication network. However, conventional transmission switching models generally represent switching actions using binary variables, without considering the availability of communication resources or whether the corresponding DC circuit breakers (DCCBs) can interrupt the fault current affected by time delay. To address these issues, this paper develops an explicit optimal info--power flow (OIPF) formulation that incorporates information transmission into the post-fault transmission-switching optimization for AC/MTDC grids. The switching status of each DC branch is determined by the availability of its communication path and associated communication resources. In addition, binary switching decisions are supplemented by a DCCB feasibility constraint that limits the estimated delayed-interruption current to the DCCB breaking capability. The resulting problem is formulated as a mixed-integer second-order cone programming (MISOCP) model that can be efficiently solved using off-the-shelf optimization solvers. Numerical results reveal that limited communication capacity and additional routing costs may render electrically beneficial switching actions infeasible or uneconomical, while delay-aware DCCB interruption constraints may favor lower-delay communication routes despite their higher costs.
\end{abstract}

\begin{IEEEkeywords}
Transmission switching, AC/multi-terminal DC grids, optimal info--power flow, communication network, DC circuit breaker
\end{IEEEkeywords}


%
\IEEEpeerreviewmaketitle

\section{Introduction}
\label{sec:I}
\subsection{Background and Motivation}
%
%
%
%
\IEEEPARstart 
The integration of multi-terminal DC (MTDC) grids based on voltage-source converters (VSCs) provides a promising solution for efficiently transmitting large-scale renewable energy from offshore generation sites to onshore AC networks. Owing to their VSC-dominated architecture, MTDC grids offer considerable operational flexibility \cite{r1}. However, AC/MTDC grids are inevitably exposed to unforeseen contingencies that may render the original transmission topology infeasible or inefficient. Under such conditions, transmission switching can reconfigure the affected system into a feasible post-contingency topology. It is therefore essential to identify a technically feasible and economically efficient reconfiguration scheme from the available candidate topologies.

\subsection{Related Works and Research Gaps}
Transmission switching, also referred to as topology reconfiguration, determines the binary on/off states of transmission lines or switches to obtain a network topology that satisfies physical operating constraints while minimizing system operating costs.

The principal computational challenge of transmission switching lies in the simultaneous determination of network topology and operating states. Each candidate switching decision alters the underlying power flow relationships and therefore requires reassessment of the corresponding system operating point. Retaining the original nonlinear power flow equations in this discrete decision process yields a mixed-integer nonlinear programming (MINLP) formulation that is prohibitively expensive to solve, particularly as the number of candidate switches increases. Existing studies, therefore, often replace the exact network power flow equations with linear approximations or convex relaxations \cite{r2,r4,r3,r23}. These formulations trade only a limited loss of power flow accuracy for a substantial gain in computational tractability. This trade-off transforms the otherwise computationally prohibitive transmission-switching problem into a form that can be effectively handled by established optimization solvers.

Extensive research has established optimization-based transmission switching frameworks for AC, DC, and integrated AC/MTDC grids \cite{r5,r6,r7,r8,r9,r10,r11}. Advances in power flow approximation, convex relaxation, and mixed-integer solution techniques have made it increasingly practical to identify economically efficient post-contingency topologies. However, identifying an optimal topology does not necessarily guarantee that the associated switching actions can be physically executed. However, major existing formulations implicitly rely on two idealized assumptions: \textit{Switching commands can always be delivered to the designated devices in a timely and reliable manner, and the corresponding switches can always successfully execute the required opening or closing actions.} These assumptions create a conceptual gap between determining an optimal topology and ensuring its physical implementation. In practice, limitations in communication and switching execution may render an otherwise optimal topology unavailable or unsafe. From this perspective, two closely related challenges must be addressed.

\textit{Can the switching command be effectively delivered?} Once an optimal post-contingency topology has been determined, the resulting switching decisions must be converted into control commands and delivered from the control center to the designated switching devices. However, such delivery cannot be taken for granted. It requires an available communication path, sufficient link capacity, and reliable information transmission within an acceptable delay. If these requirements cannot be satisfied, a switching action that is feasible from the electrical perspective may remain impossible to implement in practice. The communication network, therefore, determines not only whether a switching command can reach its destination, but also the resources required for its delivery and the time at which the corresponding action can be initiated. Consequently, more comprehensive transmission switching decision-making should explicitly account for the configuration and resource requirements of information transmission \cite{r12}.

\textit{Can the switching action be safely executed?} Even if a switching command is successfully delivered, its arrival does not necessarily guarantee that the required action can be safely completed. This issue is particularly critical for fault isolation in MTDC grids, where DC circuit breakers (DCCBs) are deployed at DC line terminals to interrupt fault currents and isolate the faulted line or protection zone \cite{r13}. Unlike routine switching operations, interrupting a DC fault is highly time-sensitive because the fault current can rise rapidly after fault inception. The actual interruption current, therefore, depends on the total delay accumulated during fault detection, command transmission, and breaker operation. In particular, a longer communication delay postpones the initiation of the DCCB action and may cause the interruption current to exceed its maximum breaking capability \cite{r14}. Therefore, whether a post-contingency topology can be safely implemented depends not only on the selected switching actions but also on the coupling among communication delay, fault-current evolution, and DCCB interruption capability. In turn, DCCB interruption limits constrain which post-contingency DC topologies can be safely established and, consequently, how power can be redistributed through the reconfigured AC/MTDC grid.

\subsection{Contributions and Outlines}
To address the two closely related challenges discussed above, this work develops an optimal info-power flow (OIPF) framework for transmission switching in AC/MTDC grids. The proposed framework jointly considers information transmission and topology reconfiguration, while explicitly capturing how time delay affects the fault current to be interrupted and, consequently, the feasibility of DCCB operation. Table \ref{tab:1} summarizes recent studies on transmission switching and highlights the two main contributions of this work\footnote{The cited studies also address other important aspects of transmission switching that fall outside the scope of this paper and are therefore omitted for brevity.}, as detailed below.
\begin{table}[t]
\caption{Comparison of representative transmission switching studies.}
\label{tab:1}
\centering
\footnotesize
\setlength{\tabcolsep}{3.0pt}
\renewcommand{\arraystretch}{1.15}

\begin{tabular}{c c cc cc c}
\toprule
\multirow{2}{*}{\textbf{Ref.}}
& \multirow{2}{*}{\textbf{Model}}
& \multicolumn{2}{c}{\textbf{Transmission Switch}}
& \multicolumn{2}{c}{\textbf{Flow}}
& \multirow{2}{*}{\shortstack{\textbf{DCCB}\\\textbf{Capability}}}
\\

\cmidrule(lr){3-4}
\cmidrule(lr){5-6}

&
&
AC Net.
&
DC Net.
&
Pow.
&
Info.
&
\\
\midrule

\cite{r5}
& MILP
& $\checkmark$
& $\times$
& $\checkmark$
& $\times$
& $\times$
\\

\cite{r6}
& MISOCP
& $\checkmark$
& $\times$
& $\checkmark$
& $\times$
& $\times$
\\

\cite{r7}
& MIQCP
& $\checkmark$
& $\times$
& $\checkmark$
& $\times$
& $\times$
\\

\cite{r8}
& MISOCP
& $\times$
& $\checkmark$
& $\checkmark$
& $\times$
& $\times$
\\

\cite{r9}
& MISOCP
& $\times$
& $\checkmark$
& $\checkmark$
& $\times$
& $\times$
\\

\cite{r10}
& MISOCP
& $\checkmark$
& $\checkmark$
& $\checkmark$
& $\times$
& $\times$
\\

\cite{r11}
& MILP
& $\times$
& $\checkmark$
& $\checkmark$
& $\times$
& $\times$
\\

\midrule

\textbf{This Work}
& MISOCP
& $\times$
& $\checkmark$
& $\checkmark$
& $\checkmark$
& $\checkmark$
\\

\bottomrule
\end{tabular}

\vspace{0.5pt}

\begin{flushleft}
\scriptsize
\ding{71} ``$\checkmark$'' indicates that the corresponding feature is considered, whereas ``$\times$'' indicates that it is not considered.
``Pow.'' and ``Info.'' denote power flow and information flow, respectively.
``MILP,'' ``MIQCP,'' and ``MISOCP'' denote mixed-integer linear programming, mixed-integer quadratically constrained programming, and mixed-integer second-order cone programming, respectively.
\end{flushleft}
\end{table}

\textbf{Explicit Optimal Info–Power Flow Formulation}: An explicit OIPF formulation is developed for transmission switching in AC/MTDC grids. Unlike conventional OPF-based transmission switching models that consider only the electrical network, the proposed formulation incorporates the communication network by introducing information-related variables. The switching status of each DC branch is coupled with the successful delivery of its control command, ensuring that a topology reconfiguration decision is implementable only when an available communication path and sufficient communication resources are provided. The resulting OIPF model is formulated as a mixed-integer second-order cone programming (MISOCP) problem and can be efficiently solved using off-the-shelf optimization solvers.
    
\textbf{DCCB-Capability-Aware Switching Formulation}: Conventional transmission switching formulations generally represent switching actions using binary variables without considering whether the corresponding DCCBs can successfully interrupt the fault current. To relax this simplified assumption, the proposed formulation additionally evaluates the feasibility of DCCB switching under time delays. Specifically, the fault current at the instant of delayed interruption is estimated and compared with the maximum breaking capability of the corresponding DCCB. A switching action is permitted only when the estimated interruption current remains within this limit. In this way, the binary switching decisions are supplemented by a practical DCCB feasibility condition, reducing the likelihood of selecting a topology that cannot be established within the considered delay.

Section~\ref{sec:2} formulates the constraints related to the electric network, while Section~\ref{sec:3} develops the constraints related to the communication network. Section~\ref{sec:4} establishes the info--power coupling constraints, and Section~\ref{sec:5} presents the corresponding optimization objective. Together, Sections~\ref{sec:2}--\ref{sec:5} constitute the complete OIPF formulation. Section~\ref{sec:6} validates the proposed model through numerical case studies, and Section~\ref{sec:7} concludes the paper.

\section{Electrical-Side Constraints}\label{sec:2}

This section presents the operating constraints associated solely with the electrical network. For ease of presentation, the main notation used is summarized as follows. $\mathcal{N}/\mathcal{E}$ denotes the system-wide node$/$branch sets. $\mathcal{V}/\mathcal{R}$ denotes the set of VSC station/RES plant.  $(\bar{\bullet})/(\underline{\bullet})$ denotes the upper$/$lower bounds of optimization variables. $(\bullet)^{ac}/(\bullet)^{dc}/(\bullet)^{vsc,v}/(\bullet)^{res,r}$ denotes the parameters or variables associated with the AC grid$/$MTDC grid$/$VSC station\#$v$$/$RES plant\#$r$. $(\bar{\bullet})/(\underline{\bullet})$ denote the upper$/$lower limits of the corresponding parameters or variables.

\subsection{Constraints of the System-Wide Power Flow}
\textit{AC grid constraints}: The AC grid constraints are summarized in Table~\ref{tab:2}. For each AC branch, $p_{ij}^{ac}$ and $q_{ij}^{ac}$ denote the power flows, while $g_{ij}^{ac}$ and $b_{ij}^{ac}$ are the branch conductance and susceptance, respectively. Using the rectangular voltage components $e_i^{ac}:=\operatorname{Re}\left\{v_i^{ac}\right\}$ and $f_i^{ac}:=\operatorname{Im}\left\{v_i^{ac}\right\}$, the auxiliary variables are defined as
$c_{ii}^{ac}:=(e_i^{ac})^2+(f_i^{ac})^2$,
$c_{ij}^{ac}:=e_i^{ac}e_j^{ac}+f_i^{ac}f_j^{ac}$, and
$s_{ij}^{ac}:=e_i^{ac}f_j^{ac}-e_j^{ac}f_i^{ac}$.
\eqref{eq:1} formulates the branch power flows, while \eqref{eq:2} approximates the apparent-power limit $\bar{s}_{ij}^{ac}$ using an $N$-sided polygon approximation \cite{r16}.
$p_i^{ac}$ and $q_i^{ac}$ denote the nodal power injections, respectively, while $G_{i,sh}^{ac}$ and $B_{i,sh}^{ac}$ denote the corresponding shunt conductance and susceptance. \eqref{eq:3} and \eqref{eq:4} impose the nodal power balance, where
$p_{i,gen}^{ac}$ and $q_{i,gen}^{ac}$ are the generator outputs,
$p_{i,load}^{ac}$ and $q_{i,load}^{ac}$ are the load demands, and
$p_{i,a2v}^{ac}$ and $q_{i,a2v}^{ac}$ represent the power transferred from the AC grid to the VSC station. For a bus without the corresponding generator, load, or VSC connection, the associated power variables are set to zero.
\eqref{eq:5} enforces the nodal voltage-magnitude limits, as $c_{ii}^{ac}=(v_i^{ac})^2$. \eqref{eq:6} enforces the generator power-output limits. Finally, \eqref{eq:7} provides the SOCP relaxation of $\left(c_{ij}^{ac}\right)^2+\left(s_{ij}^{ac}\right)^2=c_{ii}^{ac}c_{jj}^{ac}$ \cite{r4}.

\begin{table*}[t]
\centering
\caption{Constraints of the AC Grid}
\label{tab:2}
\renewcommand{\arraystretch}{0.0}
\setlength{\tabcolsep}{4pt}

\begin{tabularx}{\textwidth}{@{}>{\raggedright\arraybackslash}X >{\centering\arraybackslash}p{1.2cm}@{}}
\toprule

\multicolumn{2}{@{}l@{}}{\textit{AC branches}} \\
\addlinespace[0.3em]

$\displaystyle
p_{i j}^{a c}=g_{i j}^{a c}\left(c_{i i}^{a c}-c_{i j}^{a c}\right)+b_{i j}^{a c} s_{i j}^{a c}, \quad
q_{i j}^{a c}=-b_{i j}^{a c}\left(c_{i i}^{a c}-c_{i j}^{a c}\right)+g_{i j}^{a c} s_{i j}^{a c}, \quad
\quad \forall(i, j) \in \mathcal{E}^{a c}.
$
& \autonumlabel{eq:1} \\

$\displaystyle
-\bar{s}_{ij}^{a c} \leq \cos \left(\frac{k}{N} \pi\right) p_{ij}^{a c}+\sin \left(\frac{k}{N} \pi\right) q_{ij}^{a c} \leq \bar{s}_{ij}^{a c},
\quad \forall k \in\{1,2, \cdots, N\}, \quad \forall(i, j) \in \mathcal{E}^{a c}.
$
& \autonumlabel{eq:2} \\

\addlinespace[0.2em]
\multicolumn{2}{@{}l@{}}{\textit{AC buses}} \\
\addlinespace[0.3em]

$\displaystyle
p_i^{ac}
=
\sum_{j:(i,j)\in\mathcal{E}^{ac}}p_{ij}^{ac}
+c_{ii}^{ac}G_{i,sh}^{ac},
\quad
q_i^{ac}
=
\sum_{j:(i,j)\in\mathcal{E}^{ac}}q_{ij}^{ac}
-c_{ii}^{ac}B_{i,sh}^{ac},
\quad
\forall i\in\mathcal{N}^{ac}
$
& \autonumlabel{eq:3} \\

$\displaystyle
p_i^{a c}=p_{i, g e n}^{a c}-p_{i, l o a d}^{a c}-p_{i, a 2 v}^{a c}, \quad q_i^{a c}=q_{i, g e n}^{a c}-q_{i, l o a d}^{a c}-q_{i, a 2 v}^{a c},
\quad \forall i \in \mathcal{N}^{a c}.
$
& \autonumlabel{eq:4} \\

$\displaystyle
\left(\underline{v}_i^{a c}\right)^2 \leq c_{i i}^{a c} \leq\left(\bar{v}_i^{a c}\right)^2, \quad \forall i \in \mathcal{N}^{a c}.
$
& \autonumlabel{eq:5} \\

\addlinespace[0.2em]
\multicolumn{2}{@{}l@{}}{\textit{Generators}} \\
\addlinespace[0.3em]

$\displaystyle
\underline{p}_{i, g e n}^{a c} \leq p_{i, g e n}^{a c} \leq \bar{p}_{i, g e n}^{a c}, \quad \underline{q}_{i, g e n}^{a c} \leq q_{i, g e n}^{a c} \leq \bar{q}_{i, g e n}^{a c},
\quad \forall i \in \mathcal{N}^{a c}.
$
& \autonumlabel{eq:6} \\

\addlinespace[0.2em]
\multicolumn{2}{@{}l@{}}{\textit{SOCP relaxation}} \\
\addlinespace[0.3em]

$\displaystyle
\left(c_{i j}^{a c}\right)^2+\left(s_{i j}^{a c}\right)^2 \leq c_{i i}^{a c} c_{j j}^{a c}, \quad \forall i \in \mathcal{N}^{a c}, \quad \forall(i, j) \in \mathcal{E}^{a c}.
$
& \autonumlabel{eq:7} \\

\bottomrule
\end{tabularx}
\end{table*}

\textit{MTDC grid constraints}:
The MTDC grid constraints are summarized in Table~\ref{tab:3}. For each DC branch, $p_{jk}^{dc}$ and $p_{kj}^{dc}$ denote the power flows in opposite directions, while $r_{jk}^{dc}$ and $l_{jk}^{dc}$ represent the branch resistance and squared current magnitude, respectively. \eqref{eq:8} describes the DC branch power loss, while \eqref{eq:9} formulates the voltage drop using the squared voltage variable $u_j^{dc}:=(v_j^{dc})^2$. \eqref{eq:10} imposes the DC branch power limits. $p_j^{dc}$ denotes the nodal DC power injection, and \eqref{eq:11} establishes the nodal power balance, where $\rho^{dc}$ represents the grid polarity that $\rho^{dc}=1$ for a monopolar configuration and $\rho^{dc}=2$ for a bipolar configuration. \eqref{eq:12} enforces the squared DC voltage limits. Finally, \eqref{eq:13} provides the SOCP relaxation of $\left(p_{jk}^{dc}\right)^2=l_{jk}^{dc}u_j^{dc}$ \cite{r3}.

\begin{table*}[t]
\centering
\caption{Constraints of the MTDC Grid}
\label{tab:3}
\renewcommand{\arraystretch}{0.0}
\setlength{\tabcolsep}{4pt}

\begin{tabularx}{\textwidth}{@{}>{\raggedright\arraybackslash}X >{\centering\arraybackslash}p{1.2cm}@{}}
\toprule

\multicolumn{2}{@{}l@{}}{\textit{DC branches}} \\
\addlinespace[0.3em]

$\displaystyle
p_{j k}^{d c}+p_{k j}^{d c}=r_{j k}^{d c} l_{j k}^{d c}, \quad \forall(j,k)\in \mathcal{E}^{d c}.
$
& \autonumlabel{eq:8} \\

$\displaystyle
u_j^{d c}-u_k^{d c}=r_{j k}^{d c}\left(p_{j k}^{d c}-p_{k j}^{d c}\right), \quad \forall j,k \in \mathcal{N}^{d c}, \quad \forall(j,k)\in \mathcal{E}^{d c}.
$
& \autonumlabel{eq:9} \\

$\displaystyle
\underline{p}_{jk}^{d c} \leq p_{jk}^{d c} \leq\bar{p}_{jk}^{d c}, \quad \forall(j,k) \in \mathcal{E}^{d c}.
$
& \autonumlabel{eq:10} \\

\addlinespace[0.2em]
\multicolumn{2}{@{}l@{}}{\textit{DC buses}} \\
\addlinespace[0.3em]

$\displaystyle
p_j^{dc}
=
\rho^{dc}
\sum_{k:(j,k)\in\mathcal{E}^{dc}}p_{jk}^{dc},
\quad
\forall j\in\mathcal{N}^{dc}.
$
& \autonumlabel{eq:11} \\

$\displaystyle
\left(\underline{v}_j^{d c}\right)^2 \leq u_j^{d c} \leq\left(\bar{v}_j^{d c}\right)^2, \quad \forall j \in \mathcal{N}^{d c}.
$
& \autonumlabel{eq:12} \\

\addlinespace[0.2em]
\multicolumn{2}{@{}l@{}}{\textit{SOCP relaxation}} \\
\addlinespace[0.3em]

$\displaystyle
\left(p_{j k}^{d c}\right)^2 \leq l_{j k}^{d c} u_j^{d c}, \quad \forall j \in \mathcal{N}^{d c}, \quad \forall(j,k) \in \mathcal{E}^{d c}.
$
& \autonumlabel{eq:13} \\

\bottomrule
\end{tabularx}
\end{table*}

\textit{RES plant constraints}:
The constraints of the RES plant are summarized in Table~\ref{tab:5}. The formulation represents the entire RES plant\#$r$ as an equivalent aggregated source. \eqref{eq:14} relates the powers exchanged between the RES plant\#$r$ and the corresponding VSC station, denoted by $p_{r2v}^{res,r}$ and $q_{r2v}^{res,r}$, to the aggregated power outputs of the RES plant. \eqref{eq:15} constrains the available active power output, where $\bar{p}_{gen}^{res,r}$ denotes the maximum available renewable generation determined by natural conditions, while the apparent-power capability limit $\bar{s}_{gen}^{res}$ is approximated using an $N$-sided polygon approximation.

\begin{table*}[t]
\centering
\caption{Constraints of the RES Plant\#$r$}
\label{tab:5}
\renewcommand{\arraystretch}{0.0}
\setlength{\tabcolsep}{4pt}

\begin{tabularx}{\textwidth}
{@{}>{\raggedright\arraybackslash}X
>{\centering\arraybackslash}p{1.2cm}@{}}
\toprule

\multicolumn{2}{@{}l@{}}{\textit{RES aggregated power}} \\
\addlinespace[0.3em]

$\displaystyle
p_{r2v}^{res,r}
=
p_{gen}^{res,r},
\qquad
q_{r2v}^{res,r}
=
q_{gen}^{res,r},
\quad
\forall r\in\mathcal{R}.
$
&
\autonumlabel{eq:14}
\\

$\displaystyle
0
\le
p_{gen}^{res,r}
\le
\bar p_{gen}^{res,r},
\qquad
-\bar s_{gen}^{res,r}
\le
\cos\!\left(\frac{k\pi}{N}\right)p_{gen}^{res,r}
+
\sin\!\left(\frac{k\pi}{N}\right)q_{gen}^{res,r}
\le
\bar s_{gen}^{res,r},
\quad
\forall k\in\{1,\ldots,N\},
\quad
\forall r\in\mathcal{R}.
$
&
\autonumlabel{eq:15}
\\

\bottomrule
\end{tabularx}
\end{table*}

\begin{figure}
    \centering
    \includegraphics[width=0.8\linewidth]{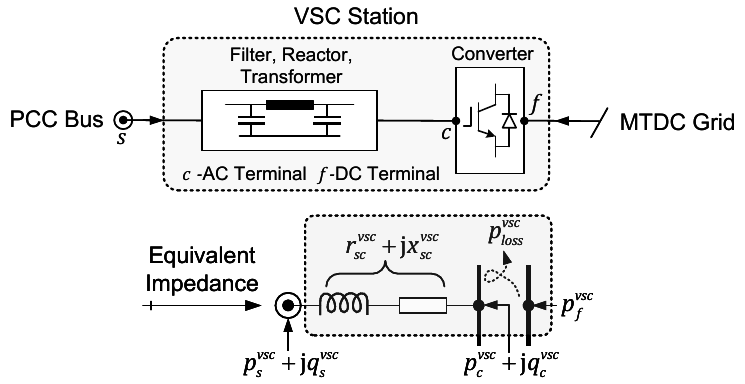}
    \caption{Equivalent impedance model of the VSC station, where the AC-side filter is omitted owing to the low harmonic content of modern modular multilevel structure.}
    \label{fig:1}
\end{figure}

\textit{VSC station constraints}:
The VSC station constraints are summarized in Table~\ref{tab:4}. As shown in Fig.~\ref{fig:1}, $p_s^{vsc,v}$ and $q_s^{vsc,v}$ denote the power injections at the PCC, while $p_c^{vsc,v}$ and $q_c^{vsc,v}$ denote those at the converter-side AC terminal. The parameters $g_{sc}^{vsc,v}$ and $b_{sc}^{vsc,v}$ are the conductance and susceptance of the branch connecting these terminals. The variables $c_{ss}^{vsc,v}$, $c_{sc}^{vsc,v}$, $s_{sc}^{vsc,v}$, $c_{cc}^{vsc,v}$, $c_{cs}^{vsc,v}$, and $s_{cs}^{vsc,v}$ are voltage-product variables analogous to those in \eqref{eq:1}. Equations \eqref{eq:16} and \eqref{eq:17} formulate the power injections at the two terminals, whose voltage magnitudes are bounded by \eqref{eq:18}. Equation \eqref{eq:19} balances the converter-side active power, VSC loss, and connected DC bus power, where $\mathcal{A}^{d2v}$ contains connection pairs $(j,v)$ between DC buses $j$ and VSC station\#$v$. The loss model in \eqref{eq:20} is quadratic in the converter current magnitude $I_c^{vsc,v}$, with $l_c^{vsc,v}$ denoting its squared magnitude and $a_{loss}^{vsc,v}$, $b_{loss}^{vsc,v}$, and $c_{loss}^{vsc,v}$ being the loss coefficients. Current bounds are imposed in \eqref{eq:21}. Equation \eqref{eq:22} couples the PCC voltage and power injections with $c_{ii}^{ac}$, $p_{i,a2v}^{ac}$, and $q_{i,a2v}^{ac}$ for $(i,v)\in\mathcal{A}^{a2v}$, the set of AC-bus--VSC-station connection pairs, or with $u_0^{res,r}$, $p_{r2v}^{res,r}$, and $q_{r2v}^{res,r}$ for $(r,v)\in\mathcal{A}^{r2v}$, the set of RES-plant--VSC-station connection pairs. Here, $u_0^{res,r}$ is the squared PCC voltage magnitude of RES plant\#$r$. Finally, the equalities $\left(c_{sc}^{vsc,v}\right)^2+\left(s_{sc}^{vsc,v}\right)^2=c_{ss}^{vsc,v}c_{cc}^{vsc,v}$, $\left(I_c^{vsc,v}\right)^2=l_c^{vsc,v}$, and $\left(p_c^{vsc,v}\right)^2+\left(q_c^{vsc,v}\right)^2=c_{cc}^{vsc,v}l_c^{vsc,v}$ are relaxed into the SOCP constraints in \eqref{eq:23}.

\begin{table*}[t]
\centering
\caption{Constraints of the VSC Station\#$v$}
\label{tab:4}
\renewcommand{\arraystretch}{0.0}
\setlength{\tabcolsep}{4pt}

\begin{tabularx}{\textwidth}{@{}>{\raggedright\arraybackslash}X >{\centering\arraybackslash}p{1.2cm}@{}}
\toprule

\multicolumn{2}{@{}l@{}}{\textit{VSC buses}} \\
\addlinespace[0.3em]

$\displaystyle
p_{s}^{vsc,v}
=
c_{ss}^{vsc,v}g_{sc}^{vsc,v}
-c_{sc}^{vsc,v}g_{sc}^{vsc,v}
+s_{sc}^{vsc,v}b_{sc}^{vsc,v},
\quad
q_{s}^{vsc,v}
=
-c_{ss}^{vsc,v}b_{sc}^{vsc,v}
+c_{sc}^{vsc,v}b_{sc}^{vsc,v}
+s_{sc}^{vsc,v}g_{sc}^{vsc,v},
\quad
s\in\mathcal{N}^{vsc,v},
\quad
(s,c)\in\mathcal{E}^{vsc,v},
\quad
\forall v\in\mathcal{V}.
$
& \autonumlabel{eq:16} \\

$\displaystyle
p_{c}^{vsc,v}
=
c_{cc}^{vsc,v}g_{cs}^{vsc,v}
-c_{cs}^{vsc,v}g_{cs}^{vsc,v}
+s_{cs}^{vsc,v}b_{cs}^{vsc,v},
\quad
q_{c}^{vsc,v}
=
-c_{cc}^{vsc,v}b_{cs}^{vsc,v}
+c_{cs}^{vsc,v}b_{cs}^{vsc,v}
+s_{cs}^{vsc,v}g_{cs}^{vsc,v},
\quad
s,c\in\mathcal{N}^{vsc,v},
\quad
(s,c)\in\mathcal{E}^{vsc,v},
\quad
\forall v\in\mathcal{V}.
$
& \autonumlabel{eq:17} \\

$\displaystyle
\left(\underline{v}_{s}^{vsc,v}\right)^2
\le
c_{ss}^{vsc,v}
\le
\left(\bar{v}_{s}^{vsc,v}\right)^2,
\quad
\left(\underline{v}_{c}^{vsc,v}\right)^2
\le
c_{cc}^{vsc,v}
\le
\left(\bar{v}_{c}^{vsc,v}\right)^2,
\quad
s,c\in\mathcal{N}^{vsc,v},
\quad
\forall v\in\mathcal{V}.
$
& \autonumlabel{eq:18} \\

\addlinespace[0.2em]
\multicolumn{2}{@{}l@{}}{\textit{VSC power loss}} \\
\addlinespace[0.3em]

$\displaystyle
p_c^{vsc,v}+p_{loss}^{vsc,v}+p_j^{dc}=0,
\quad
\forall(j,v)\in\mathcal{A}^{d2v},
\quad c\in\mathcal{N}^{vsc,v},
\quad
\forall v\in\mathcal{V}.
$
& \autonumlabel{eq:19} \\

$\displaystyle
p_{loss}^{vsc,v}
=
a_{loss}^{vsc,v}
+b_{loss}^{vsc,v}I_c^{vsc,v}
+c_{loss}^{vsc,v}l_c^{vsc,v},
\quad c\in\mathcal{N}^{vsc,v}.
$
& \autonumlabel{eq:20} \\

$\displaystyle
0\leq I_c^{vsc,v}\leq\bar{I}_c^{vsc,v},
\quad
0\leq l_c^{vsc,v}\leq(\bar{I}_c^{vsc,v})^2,
\quad c\in\mathcal{N}^{vsc,v}.
$
& \autonumlabel{eq:21} \\

\addlinespace[0.2em]
\multicolumn{2}{@{}l@{}}{\textit{Coupling relationships}} \\
\addlinespace[0.3em]

$\displaystyle
\left\{
\begin{aligned}
c_{ss}^{vsc,v} &= c_{ii}^{ac}, \quad
p_s^{vsc,v}=p_{i,a2v}^{ac}, \quad
q_s^{vsc,v}=q_{i,a2v}^{ac},
&& \forall(i,v)\in\mathcal{A}^{a2v}, \\[0.2em]
c_{ss}^{vsc,v} &= u_{0}^{res,r}, \quad
p_s^{vsc,v}=p_{r2v}^{res,r}, \quad
q_s^{vsc,v}=q_{r2v}^{res,r},
&& \forall(r,v)\in\mathcal{A}^{r2v},
\end{aligned}
\right.
\quad
s\in\mathcal{N}^{vsc,v},
\quad
\forall v\in\mathcal{V}.
$
& \autonumlabel{eq:22} \\

\addlinespace[0.2em]
\multicolumn{2}{@{}l@{}}{\textit{SOCP relaxation}} \\
\addlinespace[0.3em]

$\displaystyle
\left(c_{sc}^{vsc,v}\right)^2
+
\left(s_{sc}^{vsc,v}\right)^2
\leq
c_{ss}^{vsc,v}c_{cc}^{vsc,v},
\quad
\left(I_c^{vsc,v}\right)^2\leq l_c^{vsc,v},
\quad
\left(p_c^{vsc,v}\right)^2+\left(q_c^{vsc,v}\right)^2
\leq c_{cc}^{vsc,v}l_c^{vsc,v},
\quad
s,c\in\mathcal{N}^{vsc,v},
\quad
(s,c)\in\mathcal{E}^{vsc,v},
\quad
\forall v\in\mathcal{V}.
$
& \autonumlabel{eq:23} \\

\bottomrule
\end{tabularx}
\end{table*}

Together, constraints~\eqref{eq:1}--\eqref{eq:23} establish the system-wide AC/DC power flow constraint. These constraints ensure that the steady-state operation of the AC/MTDC grid
satisfies the aforementioned electrical operating limits, and provide the electrical foundation for the subsequent transmission switching formulation.

\subsection{Constraints for the MTDC-Side Transmission Switching}
This work explicitly considers transmission switching in the MTDC grid. Following \cite{r18}, a binary variable $\alpha_{jk}^{dc}$ is introduced to represent the connection status of each DC branch, where $\alpha_{jk}^{dc}=1$ indicates that branch $jk$ is connected, whereas $\alpha_{jk}^{dc}=0$ indicates that it is disconnected. Since $(j,k)$ and $(k,j)$ refer to the same DC branch, their switching statuses must be identical, i.e., $\alpha_{jk}^{dc}=\alpha_{kj}^{dc}$. The introduction of $\alpha_{jk}^{dc}$ renders the original DC branch model in \eqref{eq:9} inapplicable. Therefore, \eqref{eq:9} is replaced by the following mixed-integer constraints:
\begin{subequations}\label{eq:24}
\begin{equation}\label{eq:24a}
- \mathrm{M} \alpha_{jk}^{dc} \leq p_{jk}^{dc} \leq \mathrm{M} \alpha_{jk}^{dc},
\end{equation}
\begin{equation}\label{eq:24b}
(u_j^{dc}-b_{jk}^{dc})-(u_k^{dc}-t_{jk}^{dc}) = r_{jk}^{dc}(p_{jk}^{dc}-p_{kj}^{dc}),
\end{equation}
\begin{equation}\label{eq:24c}
\left(\underline{v}_j^{dc}\right)^2(1-\alpha_{jk}^{dc}) \leq b_{jk}^{dc} \leq \left(\bar{v}_j^{dc}\right)^2(1-\alpha_{jk}^{dc}),
\end{equation}
\begin{equation}\label{eq:24d}
\left(\underline{v}_k^{dc}\right)^2(1-\alpha_{jk}^{dc}) \leq t_{jk}^{dc} \leq \left(\bar{v}_k^{dc}\right)^2(1-\alpha_{jk}^{dc}),
\end{equation}
\begin{equation}\label{eq:24e}
\left(\underline{v}_j^{dc}\right)^2 \alpha_{jk}^{dc} \leq u_j^{dc}-b_{jk}^{dc} \leq \left(\bar{v}_j^{dc}\right)^2 \alpha_{jk}^{dc},
\end{equation}
\begin{equation}\label{eq:24f}
\left(\underline{v}_k^{dc}\right)^2 \alpha_{jk}^{dc} \leq u_k^{dc}-t_{jk}^{dc} \leq \left(\bar{v}_k^{dc}\right)^2 \alpha_{jk}^{dc},
\end{equation}
\end{subequations}
\[
\forall j, k \in \mathcal{N}^{dc}, \quad \forall (j, k) \in \mathcal{E}^{dc}.
\]
Where: $\mathrm{M}$ is a sufficiently large positive constant. $ b_{jk}^{dc}$ and $ t_{jk}^{dc}$ are newly introduced auxiliary variables. It can be deduced that when  $\alpha_{ij}^{dc}=1$, $u_j^{d c}-u_k^{d c}=r_{j k}^{d c}\left(p_{j k}^{d c}-p_{k j}^{d c}\right)$ is activated and otherwise inactivated.

The above transmission switching formulation can be further refined by explicitly modeling the operating states of the associated DCCBs. In practical MTDC grids, each DC branch typically has one DCCB at each terminal for protection and isolation. A branch remains in service only when both terminal breakers are closed; opening either breaker disconnects the corresponding branch. For each DC branch $(j,k)$, binary variables $\beta_{jk}^{dc}$ and $\beta_{kj}^{dc}$ are introduced to represent the operating states of the breakers located at the $j$- and $k$-side terminals, respectively. Specifically, $\beta_{jk}^{dc}=1$ indicates that the breaker adjacent to bus $j$ is closed, whereas $\beta_{jk}^{dc}=0$ indicates that it is open. The relationship between the states of terminal breakers and the branch connection status is formulated as follows:
\begin{equation}\label{eq:25}
    \alpha_{jk}^{dc} \leq \beta_{jk}^{dc}, \quad \alpha_{jk}^{dc} \leq \beta_{kj}^{dc}, \quad \alpha_{jk}^{dc} \geq \beta_{jk}^{dc}+\beta_{kj}^{dc}-1.
\end{equation}
\[
 \forall (j, k)\in \mathcal{E}^{dc}.
 \]
Where: \eqref{eq:25} establishes the logical coupling between the two terminal breakers and the corresponding DC branch status. It is the exact linear representation of
$\alpha_{jk}^{dc}=\beta_{jk}^{dc}\beta_{kj}^{dc}$.
Accordingly, $\alpha_{jk}^{dc}=1$ if and only if both terminal breakers are closed. If either $\beta_{jk}^{dc}$ or $\beta_{kj}^{dc}$ is equal to zero, the branch is disconnected and $\alpha_{jk}^{dc}=0$.

\section{Communication-Side Constraints}
\label{sec:3}
This section models the communication-side information flow. Similar to electrical power flow, information flow satisfies flow conservation. However, control commands are modeled as unsplittable commodities, each of which must be routed along a single end-to-end communication path. The superscript $(\bullet)^{info}$ denotes parameters and variables associated with information flow.

\subsection{Communication Network Architecture}
To execute transmission switching in AC/MTDC grids, each DCCB must receive a switching command from the supervisory control center through the communication network. As illustrated in Fig.~\ref{fig:1}, each DCCB is associated with a dedicated communication agent located at the corresponding VSC station, which is responsible for receiving switching commands and forwarding execution signals to the breaker. $(\bar{\bullet})/(\underline{\bullet})$ denote the upper$/$lower limits of the corresponding parameters or variables.

The communication network is represented by a directed graph $(\mathcal N^{\mathit{info}},\mathcal A^{\mathit{info}})$, where each communication agent corresponds to a communication node and each communication link is represented by two directed communication arcs. The information flow transmitted over the communication arc $(i,m)$ is denoted by $f_{im}^{\mathit{info}}$. Let $\mathcal O^{\mathit{info}}$ denote the set of communication demands. For each communication demand $o\in\mathcal O^{\mathit{info}}$, $s_o^{\mathit{info}}\in\mathcal N^{\mathit{info}}$ and $t_o^{\mathit{info}}\in\mathcal N^{\mathit{info}}$ denote its source and destination communication nodes, respectively. The binary variable $\eta_o^{\mathit{info}}\in\{0,1\}$ indicates whether communication demand $c$ is activated. Specifically, $\eta_o^{\mathit{info}}=1$ indicates that the corresponding switching command is generated and must be transmitted through the communication network, whereas $\eta_o^{\mathit{info}}=0$ indicates that no switching command is required.
\begin{figure}
    \centering
    \includegraphics[width=1\linewidth]{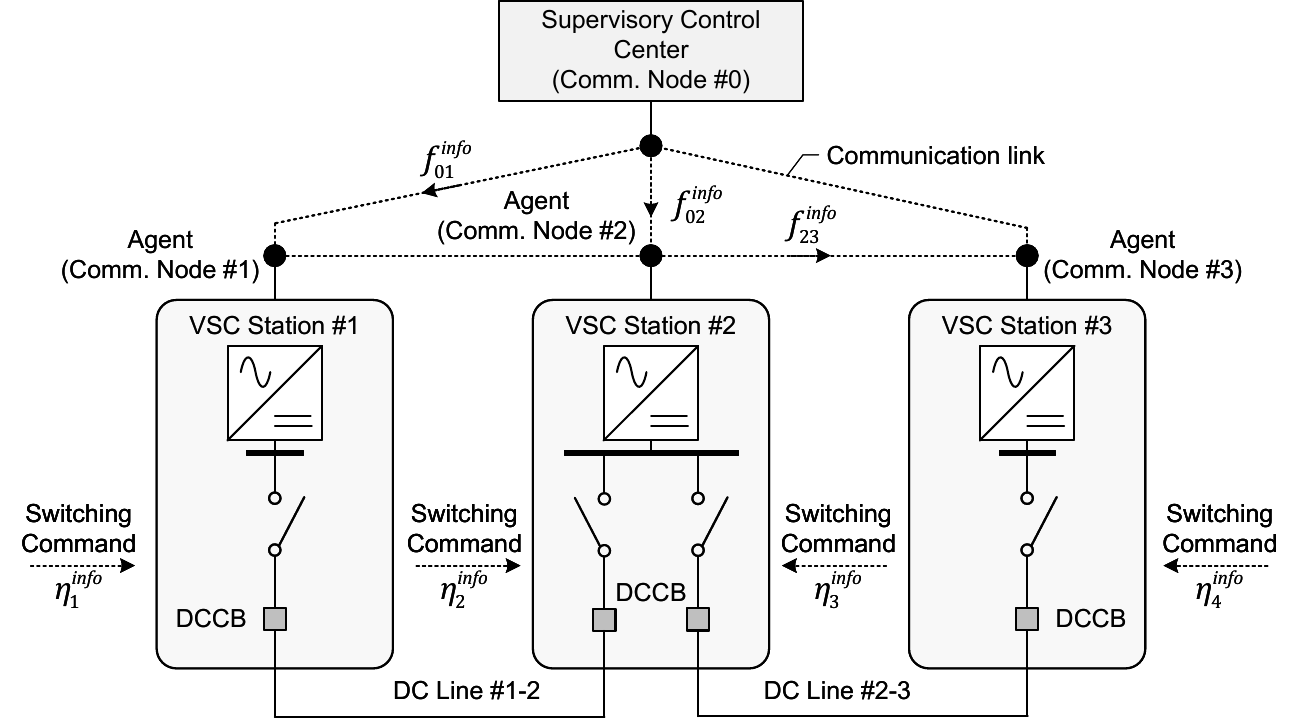}
    \caption{Schematic of the communication network architecture for AC/MTDC grids.}
    \label{fig:2}
\end{figure}

Consequently, the communication network consists of a supervisory control center, multiple communication agents, and communication links connecting them.

\subsection{Constraints for the Information Flow}
Unlike power flow, whose distribution across network branches is governed by Kirchhoff’s laws and the physical parameters of the electrical network, information flow consists of discrete communication demands that must be delivered from specified source nodes to specified destination nodes. In particular, each control command is modeled as an unsplittable communication demand and must be assigned to a complete end-to-end route \cite{r19}.

To explicitly determine the route of each activated communication demand, the binary routing variable is defined as follows:
\begin{equation}
\label{eq:26}
\begin{aligned}
z_{im,o}^{\mathit{info}}
&=
\begin{cases}
1, & \text{if demand }c\text{ is routed through arc }(i,m),\\
0, & \text{otherwise},
\end{cases}
\end{aligned}
\end{equation}
\[
\forall (i,m)\in\mathcal A^{\mathit{info}},
\quad
\forall o\in\mathcal O^{\mathit{info}}.
\]
The end-to-end routing requirement is represented by the following node balance constraint:
\begin{equation}\label{eq:27}
\begin{aligned}
&\sum_{m:(i,m)\in\mathcal{A}^{\mathit{info}}}
z_{im,o}^{\mathit{info}}
-
\sum_{m:(m,i)\in\mathcal{A}^{\mathit{info}}}
z_{mi,o}^{\mathit{info}}
\\
&\quad =
\eta_o^{\mathit{info}}
\left(
\mathbb{I}_{\{i=s_o^{\mathit{info}}\}}
-
\mathbb{I}_{\{i=t_o^{\mathit{info}}\}}
\right),
\end{aligned}
\end{equation}
\[\forall i\in\mathcal{N}^{\mathit{info}},
\quad
\forall o\in\mathcal{O}^{\mathit{info}}.\]
Where: $\mathbb{I}_{\{i=s_o^{\mathit{info}}\}}$ equals 1 if node $i$ is the source of demand $o$, and 0 otherwise. $\mathbb{I}_{\{i=t_o^{\mathit{info}}\}}$ is defined similarly for the destination. For an activated communication demand, \eqref{eq:27} requires the number of selected outgoing arcs minus incoming arcs to equal 1 at the source, $-1$ at the destination, and 0 at all intermediate nodes. For an inactive communication demand, the right-hand side is 0. Because a communication demand is treated as unsplittable, its route must not branch or be duplicated at any intermediate node. Accordingly, each communication demand is allowed to use at most one outgoing and one incoming arc at every communication node:
\begin{subequations}\label{eq:28}
\begin{align}
\sum_{m:(i,m)\in\mathcal{A}^{\mathit{info}}}
z_{im,o}^{\mathit{info}}
&\leq
\eta_o^{\mathit{info}},
\label{eq:28a}
\\
\sum_{m:(m,i)\in\mathcal{A}^{\mathit{info}}}
z_{mi,o}^{\mathit{info}}
&\leq
\eta_o^{\mathit{info}},
\label{eq:28b}
\end{align}
\end{subequations}
\[
\forall o\in\mathcal{O}^{\mathit{info}}.
\]
Where: When a communication demand is inactive, \eqref{eq:28} forces all of its associated routing variables to 0. When it is activated, at most one incoming arc and one outgoing arc can be selected at each node, thereby preventing route branching.

Equations \eqref{eq:27}--\eqref{eq:28} establish a non-branching route between the source and destination. However, neither constraint alone can guarantee that the expected communication route is selected, as illustrated in Fig.~\ref{fig:3}. To address this issue, the Miller--Tucker--Zemlin subtour elimination formulation \cite{r20} is adopted, leading to the following constraints \eqref{eq:29} and \eqref{eq:30}:
\begin{figure}
    \centering
    \includegraphics[width=1\linewidth]{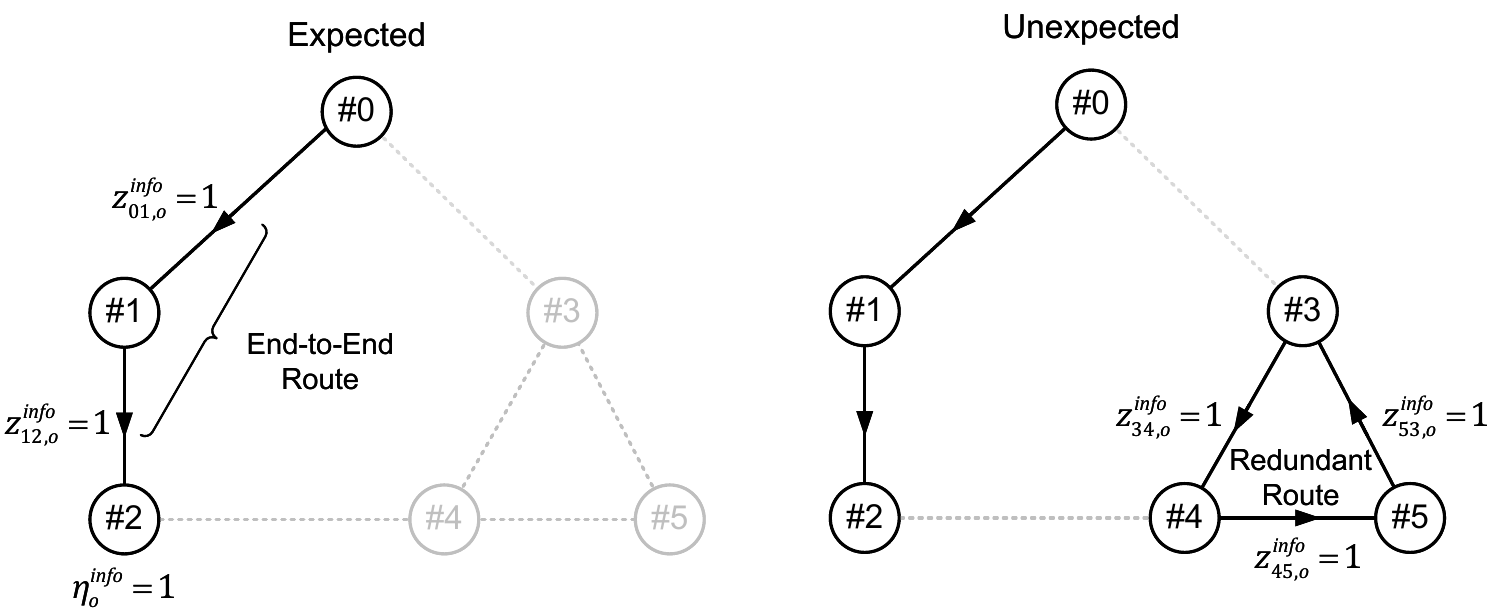}
    \caption{Illustration of undesired routing. $\#0$ is the communication source, while $\#2$ is the unique destination node. The intended communication route is $\#0\rightarrow\#1\rightarrow\#2$. However, constraints~\eqref{eq:27}--\eqref{eq:28} may also permit an undesired routing solution containing the additional directed cycle $\#3\rightarrow\#4\rightarrow\#5\rightarrow\#3$. Although this cycle does not contribute to the delivery of the communication command, the resulting routing solution still satisfies constraints~\eqref{eq:27}--\eqref{eq:28}.}
    \label{fig:3}
\end{figure}
\begin{equation}\label{eq:29}
0
\leq
l_{i,o}^{\mathit{info}}
\leq
\left(
|\mathcal{N}^{\mathit{info}}|-1
\right)
\eta_o^{\mathit{info}},
\end{equation}
\[
\forall i\in\mathcal{N}^{\mathit{info}},
\quad
\forall o\in\mathcal{O}^{\mathit{info}}.
\]
Where: $l_{i,o}^{\mathit{info}}$ is an auxiliary node-order variable indicating the sequential order in which node $i$ is traversed along the communication route of demand $o$. For instance, for the intended route $\#0\rightarrow\#1\rightarrow\#2$ in Fig.~\ref{fig:3}, the corresponding node orders can be assigned as $l_{0,o}^{\mathit{info}}=0$, $l_{1,o}^{\mathit{info}}=1$, and $l_{2,o}^{\mathit{info}}=2$.
\begin{equation}\label{eq:30}
l_{m,o}^{\mathit{info}}
\geq
l_{i,o}^{\mathit{info}}+1
-
|\mathcal{N}^{\mathit{info}}|
\left(
1-z_{im,o}^{\mathit{info}}
\right),
\end{equation}
\[
\forall i,m\in\mathcal{N}^{\mathit{info}},\quad
\forall (i,m)\in\mathcal{A}^{\mathit{info}},
\quad
\forall o\in\mathcal{O}^{\mathit{info}}.
\]
Where: When $z_{im,o}^{\mathit{info}}=1$, constraint~\eqref{eq:30} requires the node order to increase strictly from node $i$ to node $m$. For example, the undesired cycle $\#3\rightarrow\#4\rightarrow\#5\rightarrow\#3$ in Fig.~\ref{fig:3} would require $l_{4,o}^{\mathit{info}}>l_{3,o}^{\mathit{info}}$, $l_{5,o}^{\mathit{info}}>l_{4,o}^{\mathit{info}}$, and
$l_{3,o}^{\mathit{info}}>l_{5,o}^{\mathit{info}}$ simultaneously, which is impossible. Therefore, directed cycles are eliminated from the selected communication route.

$z_{im,o}^{\mathit{info}}$ specifies whether communication demand $o$ uses arc $(i,m)$, but does not directly represent the amount of bandwidth carried by that arc. Let $d_o^{\mathit{info}}$ denote the prescribed bandwidth reserved for communication demand \(o\) to support the associated DCCB control service, involving command delivery, protection data, redundancy, and etc. The aggregate information flow $f_{im}^{\mathit{info}}$  transmitted over directed communication arc $(i,m)$ is then defined as:
\begin{equation}\label{eq:info_flow_aggregation}
f_{im}^{\mathit{info}}
=
\sum_{o\in\mathcal{O}^{\mathit{info}}}
d_o^{\mathit{info}}
z_{im,o}^{\mathit{info}},
\quad
\forall (i,m)\in\mathcal{A}^{\mathit{info}}.
\end{equation}
Therefore, if multiple communication demands use the same directed arc, their bandwidth requirements are accumulated on that arc. 

Specifically, full-duplex communication with independently provisioned directional capacities is considered. The corresponding capacity constraints are:
\begin{equation}\label{eq:32}
0
\leq
f_{im}^{\mathit{info}}
\leq
\bar f_{im}^{\mathit{info}},
\quad
\forall (i,m)\in\mathcal{A}^{\mathit{info}}.
\end{equation}

Together, constraints~\eqref{eq:26}--\eqref{eq:32} establish the information flow constraint. These constraints determine a unique end-to-end communication route for each activated communication demand and aggregate the corresponding bandwidth requirements into the information flow.

\section{Info-Power-Coupling Constraints}
\label{sec:4}
This section develops the info-power coupling constraints for the proposed OIPF model. Specifically, the dependence of transmission switching actions on communication availability is explicitly modeled, while the impact of communication delay on the feasibility of DCCB interruptions is incorporated into the transmission switching formulation. $(\bar{\bullet})/(\underline{\bullet})$ denote the upper$/$lower limits of the corresponding parameters or variables.

\subsection{Constraints for the Communication Demand Activation}
Each directed DCCB switching action $(j,k)\in\mathcal{E}^{dc}$ is associated with a unique communication demand $o\in\mathcal{O}^{\mathit{info}}$. Their one-to-one correspondence is denoted by $(j,k)\Leftrightarrow o$. Therefore, the switching status transition in the power network determines whether the corresponding communication demand is activated. Let $(\hat{\bullet})$ denote the previous switching status of the corresponding breaker. The communication demand activation is formulated as:
\begin{subequations}\label{eq:33}
\begin{equation}\label{eq:33a}
\gamma_{o}^{\mathit{info}}
\geq
\beta_{jk}^{dc}-\hat{\beta}_{jk}^{dc},
\quad
\gamma_{o}^{\mathit{info}}
\geq
\hat{\beta}_{jk}^{dc}-\beta_{jk}^{dc},
\end{equation}
\begin{equation}\label{eq:33b}
\begin{gathered}
\gamma_{o}^{\mathit{info}}
\leq
\beta_{jk}^{dc}+\hat{\beta}_{jk}^{dc},
\quad
\gamma_{o}^{\mathit{info}}
\leq
2-\beta_{jk}^{dc}-\hat{\beta}_{jk}^{dc},
\end{gathered}
\end{equation}
\end{subequations}
\[
\forall (j,k)\in\mathcal{E}^{dc},
\quad
\forall o\in\mathcal{O}^{\mathit{info}},
\quad
(j,k)\Leftrightarrow o.
\]
Where: \eqref{eq:33a}--\eqref{eq:33b} enforce $\gamma_{o}^{\mathit{info}}=\left|\beta_{jk}^{dc}-\hat{\beta}_{jk}^{dc}\right|.$
Thus, $\gamma_{o}^{\mathit{info}}=1$ when the breaker status must be changed and $\gamma_{o}^{\mathit{info}}=0$ when the breaker retains its previous status.

Together, constraints \eqref{eq:33a}--\eqref{eq:33b} establish the communication-demand activation model. They identify the communication demands triggered by the optimized DCCB switching actions and establish the logical coupling between transmission switching and communication routing.

\subsection{Constraints for Delay-Aware DCCB Interruption}
The communication constraints presented above ensure that switching commands can be successfully delivered to the corresponding communication agents. However, successful command delivery does not guarantee that a DCCB can interrupt the fault current. Successful fault isolation also requires the fault current at the time of interruption to remain within the DCCB's interruption capability. Therefore, the impact of communication delay on the fault current must be considered through the following interruption capability constraints. These constraints apply only to DCCB opening actions required for fault isolation, while DCCB closing actions are excluded.

The end-to-end communication delay $T_o^{\mathit{info}}$ is obtained by accumulating the delays of all directed communication arcs contained in the selected routing path:
\begin{equation}\label{eq:34}
T_o^{\mathit{info}}
=
\sum_{(i,m)\in\mathcal A^{\mathit{info}}}
\tau_{im}^{\mathit{info}}
z_{im,o}^{\mathit{info}}, \quad
\forall o\in\mathcal O^{\mathit{info}}.
\end{equation}
Where: $\tau_{im}^{\mathit{info}}$ denotes the communication delay
associated with directed arc $(i,m)$.

The total delay $T_{jk,\mathit{tot}}^{dc}$ before the corresponding DCCB begins interrupting the fault current is formulated as:
\begin{equation}\label{eq:35}
T_{jk,\mathit{tot}}^{dc}
=
T_o^{\mathit{info}}
+
T_{jk,\mathit{fix}}^{dc},
\end{equation}
\[
\forall (j,k)\in\mathcal E^{dc},
\quad
\forall o\in\mathcal O^{\mathit{info}},
\quad
(j,k)\Leftrightarrow o.
\]
Where: $T_{jk,\mathit{fix}}^{dc}$ collects the remaining fixed delays of the associated DCCB, including fault detection, relay processing, and internal breaker operation.

The fault current to be interrupted by the corresponding DCCB depends
on the total interruption delay. To preserve the computational
tractability of the proposed MISOCP formulation, the relationship is
approximated by the following affine model, while the resulting fault
current must satisfy the DCCB breaking capability:
\begin{subequations}\label{eq:36}
\begin{align}
I_{jk,\mathit{br}}^{dc}
&:=
I_{jk,\mathit{f}}^{dc}
+
K_{jk,\mathit{f}}^{dc}
T_{jk,\mathit{tot}}^{dc},
\label{eq:37a}
\\
I_{jk,\mathit{br}}^{dc}
&\leq
\bar I_{jk,br}^{dc},
\label{eq:37b}
\end{align}
\end{subequations}
\[
\forall (j,k)\in\mathcal E^{dc},
\quad
\forall o\in\mathcal O^{\mathit{info}},
\quad
(j,k)\Leftrightarrow o.
\]
Where: $I_{jk,\mathit{br}}^{dc}$ denotes the fault current to be interrupted by the DCCB. $I_{jk,\mathit{f}}^{dc}$ denotes the fault-current value at the reference instant. $K_{jk,\mathit{f}}^{dc}$ denotes the corresponding fault-current rising-rate coefficient.

In addition to the breaking-current limit, the energy absorbed by the metal-oxide varistor (MOV) must remain within its admissible capability. Based on the analytical current-dependent energy model of the DCCB interruption process~\cite{r21}, the absorbed MOV energy and its capability constraint are formulated as:
\begin{subequations}\label{eq:37}
\begin{align}
E_{jk,\mathit{MOV}}^{dc}
&:=
\kappa_{jk}^{dc}
\left(
I_{jk,\mathit{br}}^{dc}
\right)^2,
\label{eq:39a}
\\
E_{jk,\mathit{MOV}}^{dc}
&\leq
\bar E_{jk,\mathit{MOV}}^{dc},
\label{eq:39b}
\end{align}
\end{subequations}
\[
\forall (j,k)\in\mathcal E^{dc},
\quad
\forall o\in\mathcal O^{\mathit{info}},
\quad
(j,k)\Leftrightarrow o.
\]
Where: $E_{jk,\mathit{MOV}}^{dc}$ denotes the energy absorbed by the MOV, $\kappa_{jk}^{dc}$ denotes the current-to-energy coefficient of the corresponding DCCB.

It is worth noting that DCCB operation may cause abrupt changes in the fault-current trajectory, which are not explicitly captured by
\eqref{eq:36} and \eqref{eq:37}. Hence, \(I_{jk,\mathit{f}}^{dc}\), \(K_{jk,\mathit{f}}^{dc}\), and \(\kappa_{jk}^{dc}\) should be conservatively specified such that the formulated breaking current and MOV energy are the upper bounds of their actual counterparts. Under this conservative adjustment, satisfying \eqref{eq:36} and \eqref{eq:37} provides a sufficient safety margin without explicitly modeling the terminal DCCB interruption sequence. Eventually, constraints~\eqref{eq:34}--\eqref{eq:37} establish the delay-aware DCCB interruption model by coupling the communication delay
with the DCCB breaking current and MOV energy limits.

\section{Optimization Objective}
\label{sec:5}
The proposed OIPF formulation consists of the electrical-network constraints, communication-network constraints, and electrical--communication coupling constraints developed in the
preceding subsections. Together, these constraints define the feasible solution space of the proposed optimization model. Based on this feasible region, the objective is formulated to minimize the total operating costs, including the generation cost $Cost^{st}$, the DCCB action cost $Cost^{nd}$, and the communication cost $Cost^{rd}$, such that:
\begin{equation}\label{eq:38}
\min \quad
Cost^{st}+Cost^{nd}+Cost^{rd}.
\end{equation}

The generation cost is represented by the conventional quadratic
cost function as:
\begin{equation}\label{eq:39}
Cost^{st}
:=
\sum_{i\in\mathcal{N}^{ac}}
\left[
a_{i,\mathrm{gen}}^{ac}
\left(p_{i,\mathrm{gen}}^{ac}\right)^2
+
b_{i,\mathrm{gen}}^{ac}
p_{i,\mathrm{gen}}^{ac}
+
c_{i,\mathrm{gen}}^{ac}
\right].
\end{equation}
Where: $a_{i,\mathrm{gen}}^{ac}$,
$b_{i,\mathrm{gen}}^{ac}$, and
$c_{i,\mathrm{gen}}^{ac}$ are the quadratic, linear, and constant
cost coefficients, respectively, of the generator at AC node $i$.

The DCCB action cost accounts for changes in DCCB status. Since $\gamma_{o}^{\mathit{info}}$ identifies whether the DCCB associated with communication demand $o$ changes its operating status, the action cost is formulated as:
\begin{equation}\label{eq:40}
Cost^{\mathit{nd}}
:=
\sum_{o\in\mathcal O^{\mathit{info}}}
\lambda_{o}^{\mathit{sw}}
\gamma_{o}^{\mathit{info}}.
\end{equation}
Where: $\lambda_o^{\mathit{sw}}$ denotes the operating cost associated with the DCCB switching represented by communication demand $o$.

The communication cost depends on the aggregate information flow
carried by the selected communication routes, such that:
\begin{equation}
Cost^{rd}
:=
\sum_{(i,m)\in\mathcal A^{info}}
\lambda_{im}^{info}
f_{im}^{info}.
\end{equation}
Where: $\lambda_{im}^{\mathit{info}}$ denotes the unit bandwidth cost of the communication arc $(i,m)$.

\section{Case Study}
\label{sec:6}
\begin{figure*}[!t]
\centering
\includegraphics[width=\textwidth]{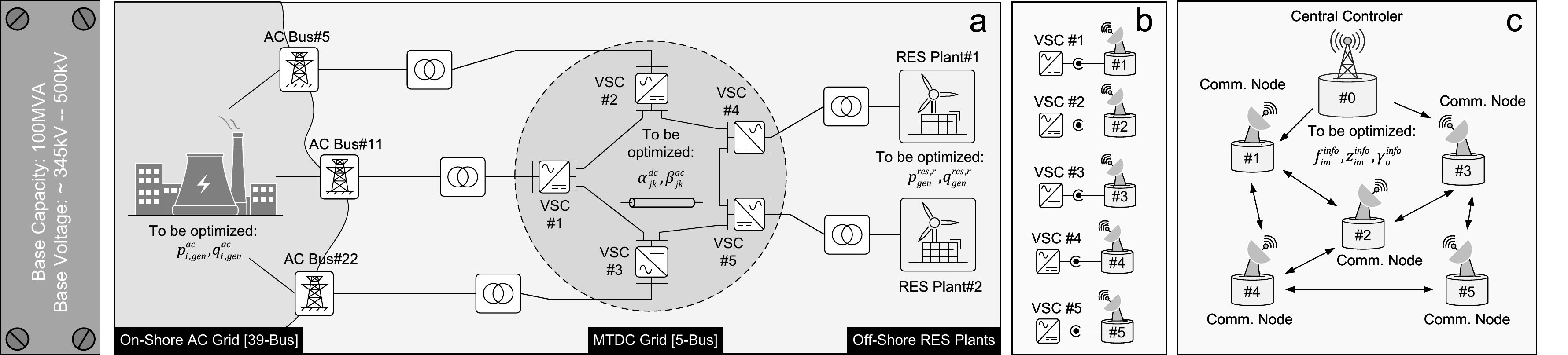}
\caption{A custom test system illustrating the delivery of offshore renewable energy to an onshore AC grid through an MTDC network. (a) presents the electrical-side system configuration. The critical variables, such as $p_{i, g e n}^{a c}$, $q_{i, g e n}^{a c}$ related to the generator power, $\alpha_{j k}^{d c}$, $\beta_{j k}^{d c}$ related to the DC topology, and $p_{g e n}^{res,r}$, $q_{g e n}^{res,r}$ related to the RES power, are optimized.  (b) shows that each VSC station is assumed to be equipped with one communication agent to receive the transmission switching command. (c) presents the communication-side system configuration. The critical variables, such as $f_{i m}^{info}$, $z_{i m}^{info}$, $\gamma_o^{info}$ related to the information flow, are optimized. }
\label{fig:4}
\end{figure*}

This section presents the case studies conducted to validate the proposed OIPF model. The custom AC/MTDC test system shown in Fig.~\ref{fig:4} is adopted for the validation\footnote{The DCCB interruption related parameters, including \(I_{jk,\mathit{f}}^{dc}\), \(K_{jk,\mathit{f}}^{dc}\), and \(\kappa_{jk}^{dc}\), can be obtained through offline EMT simulations or experimental characterization for a specific system. Since this study focuses on validating the optimization formulation, assumed parameter values are adopted in the case study to demonstrate the coupling between communication delay and DCCB interruption capability and its influence on DC transmission switching decisions.}. The key parameter settings that can potentially affect the DC transmission switching are summarized in Tables~\ref{tab:6}--\ref{tab:8}. First, the accuracy of the proposed AC/DC power flow formulation is verified by comparing the relaxed model with the original nonlinear power flow model. Subsequently, the effectiveness of the proposed OIPF formulation is demonstrated through comparisons with the conventional OPF-based transmission switching approach, highlighting the impacts of communication-related constraints and delay-aware DCCB interruption constraints on the optimal transmission switching decisions. Unless otherwise specified, all NLP models considered in the following case studies are solved using \texttt{Ipopt}\footnote{\url{https://coin-or.github.io/Ipopt/}}, whereas all MISOCP and SOCP models are solved using \texttt{Gurobi}\footnote{\url{https://www.gurobi.com/}}.

\begin{table*}[!t]
\centering
\caption{Key parameters related to the DC branches}
\label{tab:6}
\renewcommand{\arraystretch}{1.15}
\setlength{\tabcolsep}{7pt}
\footnotesize

\begin{tabular}{ccccc}
\Xhline{0.75pt}

\multirow{2}{*}{\textbf{DC Branch}}
&
\multirow{2}{*}{\textbf{Initial Status}}
&
\multirow{2}{*}{\textbf{Associated DCCBs}}
&
\multicolumn{2}{c}{\textbf{DCCB Switching}}
\\

\cline{4-5}

&
&
&
\makecell{
\textbf{Reserved Bandwidth}\\
\(\boldsymbol{d_o^{\mathit{info}}}\) \textbf{[Mb/s]}
}
&
\makecell{
\textbf{Required Action Cost}\\
\(\boldsymbol{\lambda_o^{\mathit{sw}}}\) \textbf{[\$]}
}
\\

\Xhline{0.5pt}

\(\#1-\#2\)
& Closed
& DCCB\(\#1\) \( /\) DCCB\(\#2\)
& \(5/6\)
& \(150/100\)
\\

\(\#1-\#4\)
& Closed
& DCCB\(\#3\) \( /\) DCCB\(\#4\)
& \(7/5\)
& \(250/300\)
\\

\(\#1-\#5\)
& Open
& DCCB\(\#5\) \( /\) DCCB\(\#6\)
& \(6/8\)
& \(100/120\)
\\

\(\#2-\#3\)
& Closed
& DCCB\(\#7\) \( /\) DCCB\(\#8\)
& \(4/5\)
& \(200/300\)
\\

\(\#2-\#4\)
& Open
& DCCB\(\#9\) \( /\) DCCB\(\#10\)
& \(6/7\)
& \(350/250\)
\\

\(\#2-\#5\)
& Open
& DCCB\(\#11\) \( /\) DCCB\(\#12\)
& \(5/6\)
& \(150/100\)
\\

\(\#3-\#4\)
& Open
& DCCB\(\#13\) \( /\) DCCB\(\#14\)
& \(8/4\)
& \(350/450\)
\\

\(\#3-\#5\)
& Closed
& DCCB\(\#15\) \( /\) DCCB\(\#16\)
& \(7/5\)
& \(250/350\)
\\

\(\#4-\#5\)
& Closed
& DCCB\(\#17\) \( /\) DCCB\(\#18\)
& \(6/8\)
& \(450/400\)
\\

\Xhline{0.75pt}
\end{tabular}

\vspace{0.5pt}

\begin{flushleft}
\scriptsize
\ding{71} For DC branch \(\#i-\#j\), the values on the left and right sides of \( /\) correspond to the DCCBs located at the \(\#i\)-side and \(\#j\)-side terminals, respectively.
\end{flushleft}
\end{table*}

\begin{table*}[!t]
\centering
\caption{Key parameters related to the communication nodes}
\label{tab:7}
\renewcommand{\arraystretch}{1.15}
\setlength{\tabcolsep}{12pt}
\footnotesize

\begin{tabular}{cccc}
\Xhline{0.75pt}

\textbf{Communication Node}
&
\textbf{Node Type}
&
\textbf{Associated VSC}
&
\textbf{Associated DCCBs}
\\

\Xhline{0.5pt}

\(\#0\)
& Supervisory Control Center
& --
& --
\\

\(\#1\)
& Communication Agent
& VSC\(\#1\)
& DCCB\(\#1\), DCCB\(\#3\), DCCB\(\#5\)
\\

\(\#2\)
& Communication Agent
& VSC\(\#2\)
& DCCB\(\#2\), DCCB\(\#7\), DCCB\(\#9\), DCCB\(\#11\)
\\

\(\#3\)
& Communication Agent
& VSC\(\#3\)
& DCCB\(\#8\), DCCB\(\#13\), DCCB\(\#15\)
\\

\(\#4\)
& Communication Agent
& VSC\(\#4\)
& DCCB\(\#4\), DCCB\(\#10\), DCCB\(\#14\), DCCB\(\#17\)
\\

\(\#5\)
& Communication Agent
& VSC\(\#5\)
& DCCB\(\#6\), DCCB\(\#12\), DCCB\(\#16\), DCCB\(\#18\)
\\

\Xhline{0.75pt}
\end{tabular}
\end{table*}

\begin{table*}[!t]
\centering
\caption{Key parameters of the communication links}
\label{tab:8}
\renewcommand{\arraystretch}{1.15}
\setlength{\tabcolsep}{12pt}
\footnotesize

\begin{tabular}{cccc}
\Xhline{0.75pt}

\multirow{2}{*}{\textbf{Communication Link}}
&
\multicolumn{3}{c}{\textbf{Directed Communication Arcs}}
\\

\cline{2-4}

&
\makecell{
\textbf{Available Capacity}\\
\(\boldsymbol{\bar f_{im}^{\mathit{info}}}\)
\textbf{[Mb/s]}
}
&
\makecell{
\textbf{Required Communication Cost}\\
\(\boldsymbol{\lambda_{im}^{\mathit{info}}}\)
\textbf{[\$/Mb/s]}
}
&
\makecell{
\textbf{Communication Delay}\\
\(\boldsymbol{\tau_{im}^{\mathit{info}}}\)
\textbf{[ms]}
}
\\
\noalign{\vskip 1pt}
\Xhline{0.5pt}

\(\#0-\#1\)
& \(20/24\)
& \(12/14\)
& \(1.2/1.4\)
\\

\(\#0-\#3\)
& \(20/20\)
& \(14/15\)
& \(1.0/1.2\)
\\

\(\#1-\#2\)
& \(20/18\)
& \(16/15\)
& \(0.8/0.9\)
\\

\(\#1-\#4\)
& \(25/22\)
& \(13/16\)
& \(4.0/1.0\)
\\

\(\#2-\#4\)
& \(20/16\)
& \(15/18\)
& \(0.7/0.8\)
\\

\(\#2-\#3\)
& \(20/18\)
& \(12/14\)
& \(0.9/1.0\)
\\

\(\#3-\#5\)
& \(25/20\)
& \(15/17\)
& \(1.2/1.1\)
\\

\(\#4-\#5\)
& \(20/18\)
& \(13/15\)
& \(0.8/0.9\)
\\

\Xhline{0.75pt}
\end{tabular}

\vspace{0.5pt}

\begin{flushleft}
\scriptsize
\ding{71} For communication link \(\#i-\#m\), the values on the left
and right sides of \( /\) correspond to the directed communication arcs
\(\#i\rightarrow\#m\) and \(\#m\rightarrow\#i\), respectively.
\end{flushleft}
\end{table*}

\subsection{Validation of the AC/DC Power Flow Relaxation}
We validate the accuracy of the applied power flow relaxation. Accordingly, the transmission-switching and communication-side constraints are disabled, reducing the proposed OIPF formulation to an AC/DC OPF model with a SOCP model. The benchmark is formulated using the original nonlinear AC/DC power flow equations, resulting in an NLP model. The two models are defined as follows:
\begin{itemize}
    \item OPF-SOCP: The constraints include \eqref{eq:1}--\eqref{eq:23}. The optimization objective only consider the generation cost $Cost^{st}$ formulated in \eqref{eq:39}.
    \item OPF-NLP: The OPF-NLP model adopts the original nonlinear AC/DC power flow constraints described in the manual cited in \cite{r23}. The objective function is identical to that of the SOCP-OPF model.
\end{itemize}

\begin{table}[t]
\centering
\caption{Comparison of AC/DC OPF results between NLP-based and SOCP-based Models}
\label{tab:9}

\begin{tabular}{lccc}
\Xhline{0.75pt}
\textbf{Item} & \textbf{OPF-NLP} & \textbf{OPF-SOCP}
& \textbf{Relative Error [\%]} \\
\Xhline{0.5pt}
Generation Cost [\$] & $39904.55$ & $39885.86$ & $0.0468$ \\
$P_{G1}$ [MW]  & $646.14$ & $644.50$ & $0.2538$ \\
$P_{G2}$ [MW]  & $646.00$ & $646.00$ & $0.0000$ \\
$P_{G3}$ [MW]  & $643.85$ & $643.49$ & $0.0559$ \\
$P_{G4}$ [MW]  & $631.91$ & $631.89$ & $0.0032$ \\
$P_{G5}$ [MW]  & $508.00$ & $508.00$ & $0.0000$ \\
$P_{G6}$ [MW]  & $636.67$ & $636.64$ & $0.0047$ \\
$P_{G7}$ [MW]  & $580.00$ & $580.00$ & $0.0000$ \\
$P_{G8}$ [MW]  & $564.00$ & $564.00$ & $0.0000$ \\
$P_{G9}$ [MW]  & $631.20$ & $632.74$ & $0.2440$ \\
$P_{G10}$ [MW] & $663.86$ & $663.00$ & $0.1295$ \\
\Xhline{0.75pt}
\end{tabular}
\end{table}

Table~\ref{tab:9} compares the AC/DC OPF solutions obtained from the OPF-NLP model and the OPF-SOCP model. It can be observed that the total generation costs obtained by the two models are highly consistent. Furthermore, the active power outputs of all generators exhibit only minor deviations. These results demonstrate that the proposed SOCP relaxation introduces only negligible approximation errors to the original nonlinear AC/DC power flow model while significantly improving the convexity of the optimization problem. Therefore, the proposed SOCP formulation provides an acceptable accurate and computationally tractable foundation for the subsequent OIPF-based transmission switching decisions.

\subsection{Comparison of DC Transmission Switching}
We demonstrate the impact of information flow on transmission switching decisions. The following models are considered, all of which are formulated as MISOCP problems:

\begin{itemize}
    \item OPF: The constraints include \eqref{eq:1}--\eqref{eq:8}, \eqref{eq:10}--\eqref{eq:25}, \eqref{eq:33}. The optimization objective is set as $Cost^{st}+Cost^{nd}$, and the communication cost is not involved.
    \item OIPF-a: The constraints include \eqref{eq:1}--\eqref{eq:8}, \eqref{eq:10}--\eqref{eq:33}, while the delay-aware DCCB interruption constraints in \eqref{eq:34}--\eqref{eq:37} are excluded. The objective function is set to \eqref{eq:38}.
    \item OIPF-b: The constraints includes \eqref{eq:1}--\eqref{eq:8} and \eqref{eq:10}--\eqref{eq:37}. The objective function is set to \eqref{eq:38}. It represents the complete OIPF model.
\end{itemize}

\begin{figure}
    \centering
    \includegraphics[width=1\linewidth]{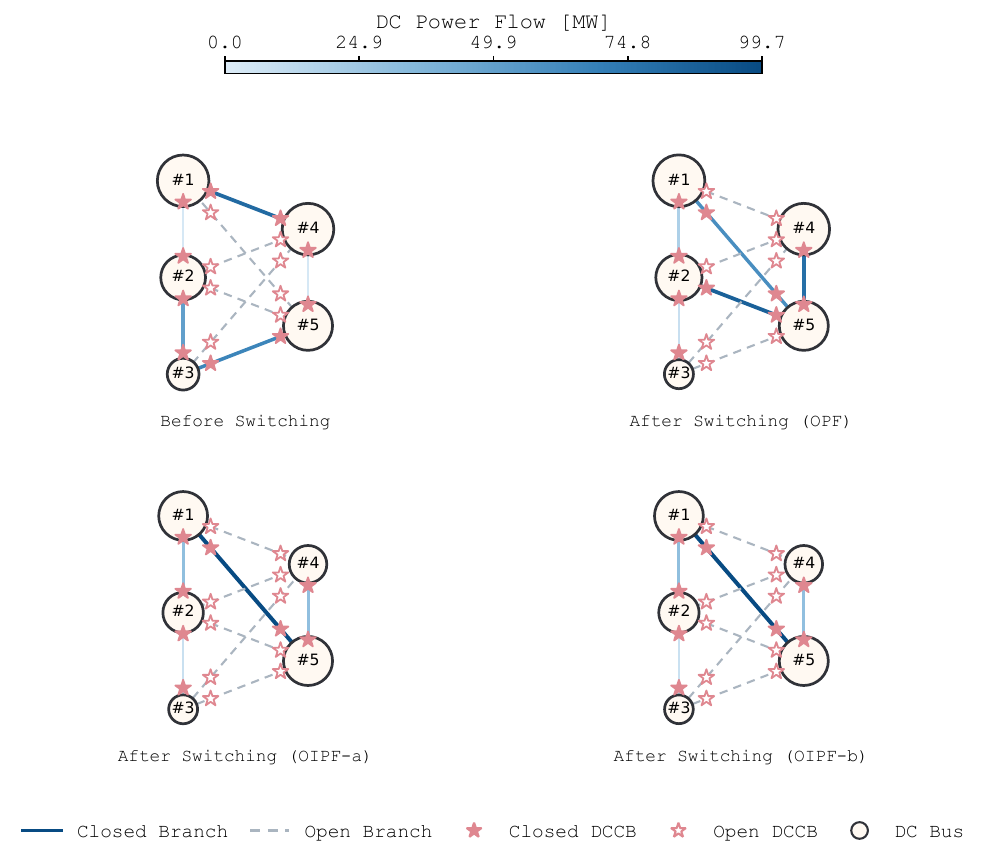}
    \caption{DC transmission switching results. Darker (lighter) branch colors indicate larger (smaller) power flow magnitudes, and larger (smaller) node sizes represent higher (lower) power-injection levels. The plotted branch power flow magnitude is $(|p_{jk}^{dc}|+|p_{kj}^{dc}|)/2$.}
    \label{fig:5}
\end{figure}

The initially in-service DC branches \(\#1-\#4\) and \(\#3-\#5\) are assumed to experience faults and must be disconnected. Fig.~\ref{fig:5} presents the resulting transmission switching decisions obtained using the different models. As observed, OPF additionally closes DC branch \(\#2-\#5\), whereas OIPF-a and OIPF-b retain this branch in its open state. This difference can be interpreted according to the parameters provided in Tables \ref{tab:5} and \ref{tab:8}. Closing this branch requires additional control commands at DCCB\(\#11\) and DCCB\(\#12\). The additional DCCB commands would exceed the limited communication capacity. Under OPF, closing branch \(\#2-\#5\) incurs only DCCB action cost of $\$150+\$100=\$250$. In contrast, OIPF-a and OIPF-b also account for the communication cost of delivering the corresponding DCCB control commands. From the resulting decisions, we can deduce that the generation cost savings achieved by closing branch \(\#2-\#5\) are insufficient to offset the associated increase in communication cost. Therefore, OIPF-a and OIPF-b retain branch \(\#2-\#5\) in its open state.

\subsection{Comparison of Communication Routing}
\begin{figure}
    \centering
    \includegraphics[width=.9\linewidth]{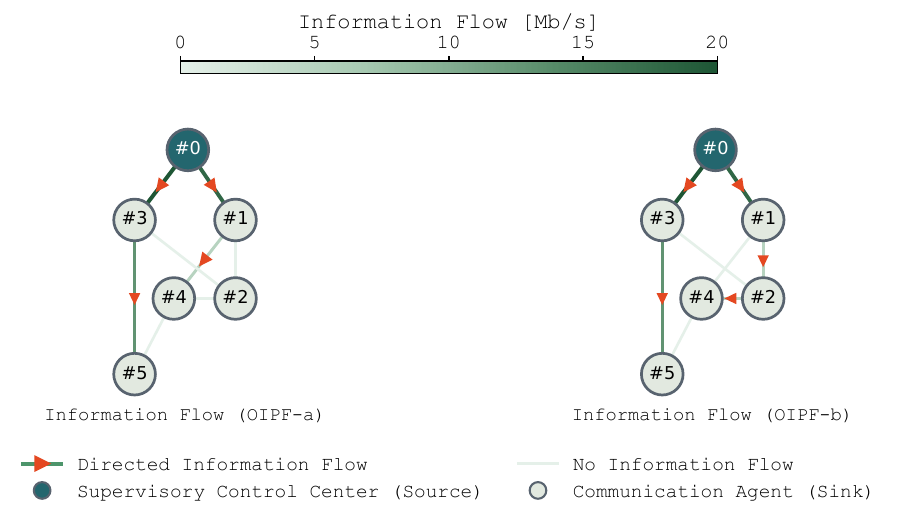}
    \caption{Communication routing results. Darker (lighter) link colors indicate larger (smaller) information-flow magnitudes, and larger (smaller) node sizes represent higher (lower) aggregate bandwidth requirements for the associated DCCB control commands. Arrows indicate the directions of information flows.}
    \label{fig:6}
\end{figure}

Although OIPF-a and OIPF-b yield similar DC transmission switching decisions, their communication routing decisions differ substantially, as illustrated in Fig.~\ref{fig:6}. It can be observed that, to deliver the DCCB control command associated with sink \(\#4\), OIPF-b selects the combined arcs \(\#1\rightarrow\#2\rightarrow\#4\), rather than the direct arc \(\#1\rightarrow\#4\) selected by OIPF-a. From a purely communication-cost perspective, this routing decision is less economical. The direct arc has a unit communication cost of \(\$13/\mathrm{Mb/s}\), whereas the combined arcs have an aggregated unit cost of \(\$16/\mathrm{Mb/s}+\$15/\mathrm{Mb/s}=\$31/\mathrm{Mb/s}\). However, the direct communication arc \(\#1\rightarrow\#4\) introduces a relatively long delay, postponing command delivery and potentially causing the fault current at interruption to exceed the DCCB's breaking capability. Consequently, the required breaking current and MOV energy can exceed the DCCB's corresponding interruption limits. Therefore, OIPF-b accepts a higher communication cost to ensure a lower-delay communication route, thereby ensuring that the associated DCCB has sufficient capability to safely interrupt the fault current.

\section{Conclusion}
\label{sec:7}
This paper developed a DCCB-capability-aware OIPF model for the DC transmission switching in AC/MTDC grids. The key insight is that conventional OPF-based transmission switching has inherent limitations: an electrically feasible switching action may not be physically executable due to command-delivery constraints associated with DCCB actions. The case studies reveal the importance of accounting for info-power coupling and DCCB interruption capability in transmission switching decisions. Overall, the proposed OIPF model represents a step toward obtaining transmission switching decisions that are more practically implementable and that will be needed with the planned installation of DC switching stations inside the HVDC grid.
\ifCLASSOPTIONcaptionsoff
  \newpage
\fi



%
\bibliographystyle{IEEEtran}
\bibliography{ref}

%








\end{document}